\documentclass[letterpaper,10pt,conference]{ieeeconf}
\IEEEoverridecommandlockouts
\usepackage{booktabs}
\usepackage{tabularx}
\usepackage{array}
\usepackage{cite}
\usepackage{amsmath,amssymb,amsfonts}
\usepackage{algorithmic}
\usepackage{algorithm}
\usepackage{graphicx}
\usepackage{booktabs}
\usepackage{multirow}
\usepackage{tabularx}
\usepackage{array}
\usepackage{makecell}
\usepackage{textcomp}
\usepackage{xcolor}
\usepackage{hyperref}
\usepackage{microtype}
\begin{document}

\title{A Two-Stage Reflection and Reprompting Framework for LLM-Based Solution of Petri Net Reachability Problems in Industrial Applications\\
}

\author{
\begin{tabular}{cc}
\begin{tabular}[t]{@{}c@{}}
\textbf{Ruimin Hu}\\
\textit{Department of Chemical Engineering}\\
\textit{Imperial College London}\\
London, UK\\
ruimin.hu25@imperial.ac.uk
\end{tabular}
&
\begin{tabular}[t]{@{}c@{}}
\textbf{Mehmet Mercang{\"o}z}\\
\textit{Department of Chemical Engineering}\\
\textit{Imperial College London}\\
London, UK\\
m.mercangoz@imperial.ac.uk
\end{tabular}
\end{tabular}
}

\maketitle

\begin{abstract}
Manufacturing systems exhibit strong concurrency, synchronization, and contention for shared reusable resources, which makes fast and reliable scheduling and verification challenging. Petri nets provide a rigorous formalism for modeling such discrete-event manufacturing systems, but reachability analysis and solving remain difficult for conventional graph search or optimization-based solvers, particularly under state-space explosion and evolving production requirements. Recently, Large language models (LLMs) have shown promise as flexible planners that can generate candidate action sequences from textual specifications. However, direct use of LLMs for Petri net reachability remains unreliable. This paper proposes an LLM-based solving framework augmented with a two-stage reflection and reprompting mechanism. The combined effects of reflection and re-clarification improve the accuracy of feasible sequence generation. The proposed method is evaluated on an industrial case modeled as a Petri net. Under a fixed Petri net structure, the proposed strategy is assessed on six solvable reachability configurations. The results demonstrate improved reliability and stability in solving Petri net reachability problems. The proposed framework is further evaluated across multiple LLMs, which indicates that the framework is not tied to any specific model.
\end{abstract}


\section{Introduction}
With the advance of Industry 4.0 and intelligent manufacturing, manufacturing systems are increasingly shifting toward high-mix, low-volume operations \cite{rossit2019industry}. Market uncertainty and customization requirements often lead to dynamic order arrivals, parallel processing stages, and contention for shared resources. As a result, a central challenge in manufacturing is to achieve fast and effective scheduling, which underpins production line design and verification, process route evaluation, and rapid iteration of operational strategies.

Petri nets provide a rigorous formalism for representing the behavior of discrete manufacturing systems, including sequencing, concurrency, and synchronization \cite{murata2002petri}. Consequently, Petri nets have been widely adopted for modeling and analyzing discrete-event systems. In typical formulations, operations and events are modeled as transitions, whereas buffers, resources, and system states are represented by places. Resource capacities can be encoded via token counts, arc weights, or explicit capacity constraints \cite{zurawski1994petri}. Petri nets therefore offer a clear formal description of system structure while enabling characterization of system dynamics as state evolution driven by transition firings. In addition, the modular and scalable structure of Petri nets supports efficient model construction for large-scale and complex manufacturing systems.

Although Petri nets provide a rigorous formalism for modeling manufacturing processes and representing system states, solving Petri net reachability and synthesizing valid firing sequences remain challenging in practice. Conventional approaches typically depend on explicit construction and exploration of the reachability graph. In many implementations, the state space is traversed using generic graph-search strategies such as depth first search (DFS) and breadth first search (BFS) \cite{chu1997deadlock}. Optimization-based formulations, including integer linear programming and constraint programming, have also been investigated to derive feasible firing sequences. Despite the ability of these methods to provide exact solutions when computational resources permit, they often suffer from state-space explosion as the Petri net increases in scale and concurrency, which results in substantial computational and memory overhead. Heuristic approaches can produce solutions with lower latency, but they often rely on carefully engineered and model-specific rules. When the net structure or production requirements change, autonomous redesign of these heuristics is frequently required \cite{baldea2025automated}. These limitations motivate the development of more adaptive and model-agnostic alternatives.

In recent years, large language models (LLMs) have demonstrated strong capabilities in planning and reasoning \cite{gill2025leveraging} and have increasingly been discussed as a promising route toward artificial general intelligence (AGI) \cite{augenstein2024factuality}. This progress suggests a new direction for Petri net reachability analysis and firing-sequence synthesis: instead of explicitly constructing the reachability graph or relying on expertise-intensive handcrafted heuristics, an LLM can generate candidate firing sequences directly from a text-encoded model specification and goal constraints. This paradigm has practical appeal in terms of rapid adaptation and deployment, particularly in manufacturing settings in which system configurations and production requirements change frequently.

However, direct use of an LLM as a solver for Petri net reachability remains challenging. A generated firing sequence may include transitions that are not enabled under the current marking. Even when each step is locally enabled, execution can still enter deadlock or other nonprogressing states, or terminate at a marking that does not satisfy the target specification. In addition, LLMs may produce incorrect reachability judgments, which can lead to false acceptance of infeasible goals or false rejection of feasible ones. These failure modes are particularly pronounced in systems with shared reusable resources, where intermediate reasoning may overlook transient resource availability and thereby produce sequences that fail during verification. Therefore, practical viability of LLM-driven Petri net solving requires reliable verification mechanisms and effective guidance strategies to improve correctness and robustness.

To address the aforementioned challenges of using LLMs for Petri net solving, this paper proposes an LLM-driven two stage reflection and reprompting framework and evaluates the framework on an industrial case study. The LLM first generates a candidate firing sequence. A Petri net simulator then replays the sequence step by step, strictly checking transition enabledness at each marking and determining whether the target specification is satisfied. When verification fails, a two-stage correction procedure is applied instead of directly requesting another unconstrained full output. In the reflection stage, the model is restricted to output only a structured report that diagnoses the failure mode, pinpoints the first erroneous step, summarizes the current marking and the set of enabled transitions, and proposes a repair plan together with avoidance patterns. In the second stage, the model generates a revised firing sequence conditioned on the reflection report and verifier feedback. By enforcing closed-loop verification while providing structured diagnostic guidance, the proposed mechanism improves the reliability of LLM-generated solutions and enhances the ability of the model to satisfy constraints inherent in Petri net representations of manufacturing scenarios.

The main contributions of this paper are summarized as follows:
\begin{itemize}
  \item We propose and implement a closed-loop, LLM-driven solving framework for reachability solving of Petri nets and firing-sequence synthesis.
  \item We design a two-stage reflection and reprompting mechanism that structures feedback on failures and guides the model to perform strategic repairs, which improves solution stability under constraints.
  \item We conduct an experimental evaluation on an industrial case study to validate the effectiveness of the proposed approach.
\end{itemize}

The remainder of this paper is organized as follows. Section~\ref{sec.B} reviews Petri nets and industrial applications of LLMs. Section~\ref{sec.C} presents the proposed two-stage reflection and reprompting approach in detail. Section~\ref{sec.D} describes the experimental setup and the industrial case study. Section~\ref{sec.E} reports and analyzes the experimental results. Finally, Section~\ref{sec.F} concludes the paper and discusses directions for future research.

\section{Background}
\label{sec.B}

\subsection{Petri Nets Preliminaries}

Petri nets provide a formal and graphical modeling framework for discrete event systems with concurrency, synchronization, and resource sharing. In this paper, we consider a place and transition Petri net defined as:

\begin{equation}
\mathcal{N}=(P,T,Pre,Post)
\end{equation}

where $P$ is a finite set of places, $T$ is a finite set of transitions with $P\cap T=\emptyset$, and $Pre,Post: P\times T \rightarrow \mathbb{N}_0$ denote the input and output arc weight functions.
An arc $(p,t)$ exists iff $Pre(p,t)>0$, and an arc $(t,p)$ exists iff $Post(p,t)>0$.

A marking is a function $M:P\rightarrow \mathbb{N}$, where $M(p)$ is the number of tokens in place $p$.
A transition $t\in T$ is enabled at marking $M$ if

\begin{equation}
\forall p\in P,\quad M(p)\ge Pre(p,t)
\end{equation}

If $t$ is enabled, firing $t$ yields a new marking $M'$ defined by
\begin{equation}
\forall p\in P,\quad M'(p)=M(p)-Pre(p,t)+Post(p,t).
\label{eq:firing}
\end{equation}

Given an initial marking $M_0$ and a target marking $M_g$,
$M_g$ is reachable from $M_0$ in $\mathcal{N}$ if there exists a finite firing sequence
$\sigma=(t_1,t_2,\ldots,t_k)$ such that
\begin{equation}
M_0 \xrightarrow{t_1} M_1 \xrightarrow{t_2} \cdots \xrightarrow{t_k} M_k = M_g.
\label{eq:reach}
\end{equation}

\subsection{Petri Nets in Industrial Systems}

The concurrency inherent in industrial manufacturing systems aligns well with the expressive power of Petri nets, which has motivated broad adoption of Petri nets for modeling production processes and supporting resource decision-making \cite{Zurawski1994,tuncel2007applications}. In industrial applications, Petri nets provide a graphical yet formal representation of system behavior, which enables analysis and supervision of state evolution, particularly for deadlock prevention \cite{Mejia2018}. Moreover, Petri net models support both qualitative and quantitative assessment, ranging from structural and behavioral properties to performance evaluation when appropriate extensions are employed \cite{zhou1998modeling}. Finally, Petri nets can be integrated with optimization and control algorithms and can serve as a modeling backbone for systematic improvement of manufacturing operations \cite{DooYongLee1994}.

A substantial body of work leverages Petri nets to model and optimize industrial production systems. For instance, in the Job Shop Scheduling Problem (JSSP), Lassoued et al. propose a reinforcement learning-based approach to derive scheduling policies \cite{lassoued2024introducing}. For flexible manufacturing systems (FMS), Hu et al. model system dynamics using Petri nets and combine deep reinforcement learning with a graph convolutional network (GCN) to address the resulting decision problem \cite{hu2020petri}. Similarly, \cite{luo2024petri} integrates reinforcement learning with Petri net modeling to address real-time scheduling in automated manufacturing systems. Beyond reinforcement learning, Petri net models are often coupled with metaheuristic optimization methods \cite{luo2018deadlock,han2018petri}. Another widely used class of methods comprises tree search-based approaches, which typically rely on handcrafted heuristics to guide the search toward promising scheduling decisions \cite{mejia2017petri,huang2012scheduling}. A related line of work optimizes the configuration of the Petri net model itself rather than only the firing decisions, da Silva et al. iteratively configure a T-timed coloured Petri net through a mixed-integer linear programming (MILP) model to plan rail freight volumes and rolling stock \cite{dasilva2025iterative}. However, many of these methods depend heavily on expert-designed modeling choices and algorithm-specific heuristics, which require substantial domain knowledge. When a manufacturing system changes because of disruptions or unexpected events, performance may degrade and redesign of the solution strategy is often required.

\subsection{LLMs for Industrial Planning}

In recent years, LLMs have attracted growing attention in intelligent manufacturing. A key advantage of LLMs is the ability to extract actionable knowledge representations from natural language, structured data, and process rules, and to generate decision recommendations or executable action sequences accordingly \cite{ren2025industrial}. Integration enabled by LLMs can facilitate process planning and manufacturing workflow optimization, which can reduce the time and cost associated with manual planning and improve operational efficiency \cite{ni2025large}. Despite this potential, manufacturing involves complex and domain specific knowledge, which poses substantial challenges to effective and reliable use of LLMs.

Prior work has explored LLMs for manufacturing related decision support and planning. Xia et al. propose an error-guided fine-tuning strategy that strengthens the capability of LLMs to answer manufacturing specific queries and to generate valid domain-specific code \cite{xia2024leveraging}. Song et al. develop LLM-Planner, which leverages LLMs to perform few-shot planning for embodied or software agents \cite{song2023llm}. Zheng et al. introduce a hierarchical approach that incrementally decomposes a complex objective into a task tree, which reduces planning burden at each step by combining human expert knowledge with guidance generated by LLMs \cite{zhen2023robot}. Wang et al. develop an LLM-based multi-agent system for the flexible job-shop scheduling problem \cite{wang2025masc}. Reflection refers to the ability of an LLM to inspect the reasoning of itself or other agents, assess validity, identify errors, and adjust subsequent actions based on this assessment to obtain more accurate outcomes. Building on this idea, Cao et al. propose a hierarchical reflection scheme for scheduling \cite{cao2025reflecsched}.

Overall, existing studies indicate that LLMs can serve as powerful planners or decision assistants in manufacturing. However, most manufacturing planning problems are governed by strict constraints, under which plans generated by LLMs may still include infeasible steps or terminate at outcomes that do not match the goal specification. In this work, Petri nets are adopted as the modeling backbone, and a verifier in the loop LLM solving framework is developed to generate and repair firing sequences toward exact target markings.

\section{Methodology}
\label{sec.C}

To mitigate the tendency of LLMs to produce hallucinated or invalid steps when Petri net solving is performed directly, which often prevents generation of enabled and goal-reaching firing sequences, this paper proposes an LLM-driven closed-loop solving framework augmented with a two-stage reflection and reprompting mechanism. The key idea is to let the LLM generate candidate firing sequences, replay and verify each candidate step by step using a Petri net simulator to check transition enabledness, provide structured diagnostic feedback when verification fails, and iteratively repair the candidate until a feasible sequence that satisfies target marking constraints is obtained.

\subsection{Method Overview}

Given a Petri net structure together with initial and target markings, the model is converted into a standardized textual specification and provided to the LLM using a predefined prompt template, such that a candidate firing sequence is returned in a fixed format. This specification is a JSON object. Its structure field lists the input and output places and arc weights of each transition, a direct transcription of $Pre$ and $Post$, as in \texttt{"T1": \{"in": \{"P1":1,"P2":1\}, "out": \{"P3":1\}\}}. The markings field gives $M_0$ and $M_g$ as place--token pairs, and any place not listed is taken to be empty. Because places, transitions, arcs, and markings are all encoded explicitly, the encoding is lossless and applies to a place/transition net of any size. The candidate sequence is then replayed step by step using a Petri net simulator. At each step, the verifier checks whether the selected transition is enabled under the current marking. If any transition is not enabled, such as, the transition cannot be fired, the sequence is deemed invalid and execution is terminated immediately. The verifier returns diagnostic information, and the procedure enters the reflection and reprompting stage. If the sequence can be executed to completion, the resulting marking is further examined to determine whether terminal constraints are satisfied.

\begin{figure}[htbp]
  \centering
  \includegraphics[width=\linewidth]{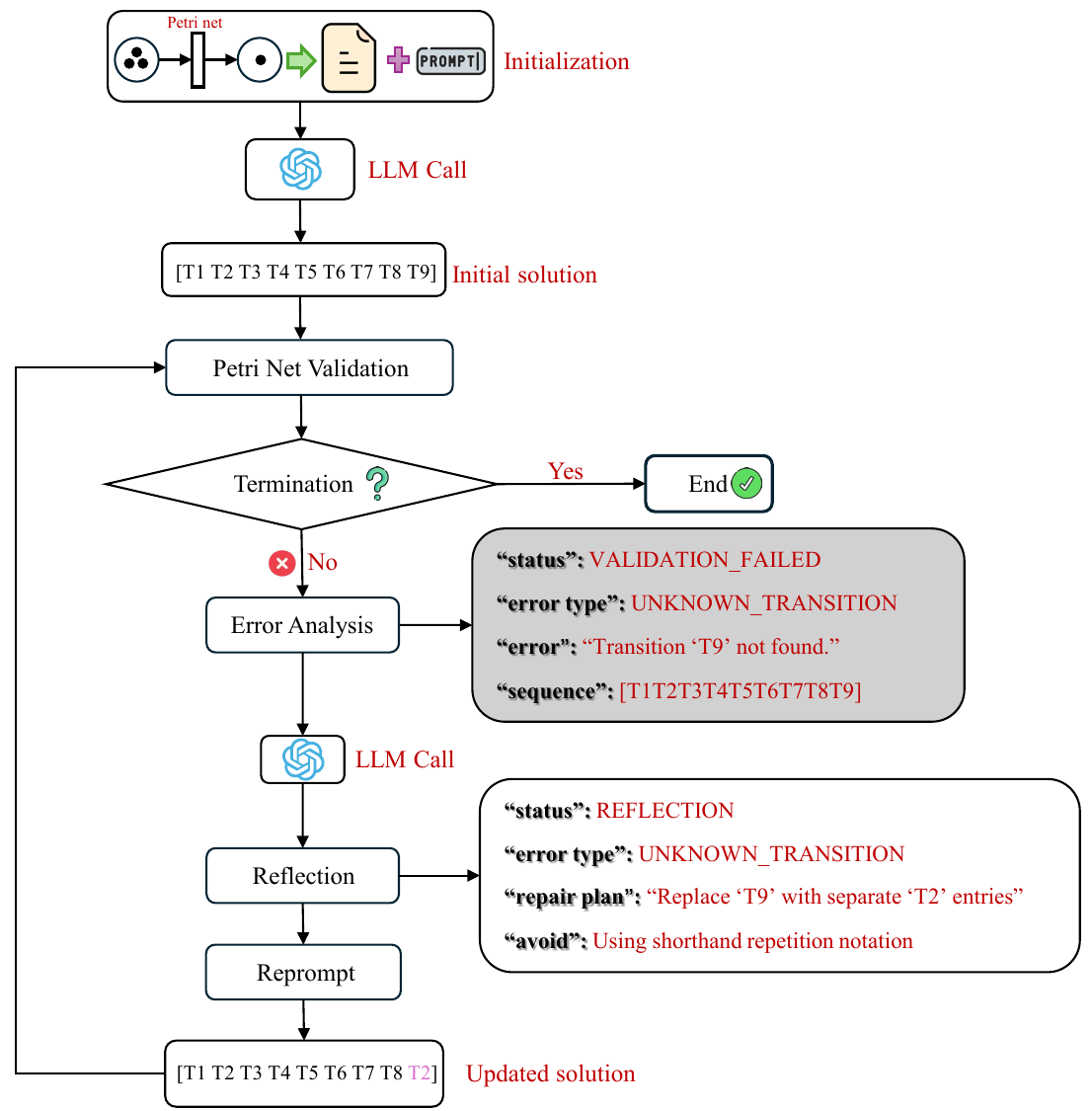}
  \caption{Overall framework of two-stage reflection and reprompting for Petri net solving.}
  \label{fig1}
\end{figure}

When verification fails, repeated sampling without guidance is not adopted. Instead, a two-stage reflection and reprompting strategy is employed. Key error signals, including failure type and related diagnostic information, are provided to the LLM. In the reflection stage, the model analyzes the failure, produces an actionable repair plan, and summarizes error patterns to avoid. In the subsequent stage, the reflection output is combined with the original problem specification to construct an updated prompt, which guides the LLM to generate a revised firing sequence. The overall framework is illustrated in Fig.~\ref{fig1}.

\subsection{Reflection and Reprompting Strategy}

To improve reliability of LLM-based Petri net solving, a reflection and reprompting strategy is proposed. The core objective is to convert verification failures associated with a candidate firing sequence into structured and actionable diagnostic feedback. This feedback first prompts the model to analyze the failure, attribute the failure to a concrete cause, and formulate a repair plan, and then guides regeneration of a revised firing sequence. In this way, the strategy reduces inefficiency and instability that arise from repeated sampling without guidance.

\begin{enumerate}
  \item \textbf{Structured diagnostic feedback:}
  During step by step replay of a candidate sequence, verification failures are
  categorized into two typical types: (i) \emph{intermediate step violations},
  where a chosen transition is ill formed, undefined, or not enabled under
  the current marking; and (ii) \emph{terminal goal violations}, in which the final
  marking fails to satisfy the target specification. The verifier returns
  structured diagnostic fields to support subsequent prompt construction. Representative
  error signals include formatting errors, undefined transitions, not-enabled
  transitions, and non-progressing transition loops.

  \item \textbf{Reflection stage:}
  Once a failure is detected, the procedure enters the Reflection stage, which
  consumes the structured diagnostics described above. This stage is not intended
  to directly generate a new sequence. Instead, it asks the model to perform
  failure attribution and strategy summarization, identify which class of
  constraint conflict caused the failure, determine how the sequence should be
  repaired, and extract patterns to avoid, thereby reducing repeated mistakes in
  later iterations.

  \item \textbf{Reprompting stage:}
  After obtaining reflection outcomes, the method proceeds to the  reprompting
  stage. In this stage, the reflection conclusions, the original problem specification,
  and constraints implied by the current failure state are integrated into an updated
  prompt, which guides the LLM to produce a new candidate firing sequence. Compared
  with direct resampling, reprompting improves efficiency of LLM-based solving and
  reduces probability of recurring errors.
\end{enumerate}

\section{Illustrative Case Study}
\label{sec.D}

\begin{figure}[htbp]
  \centering
  \includegraphics[width=\linewidth]{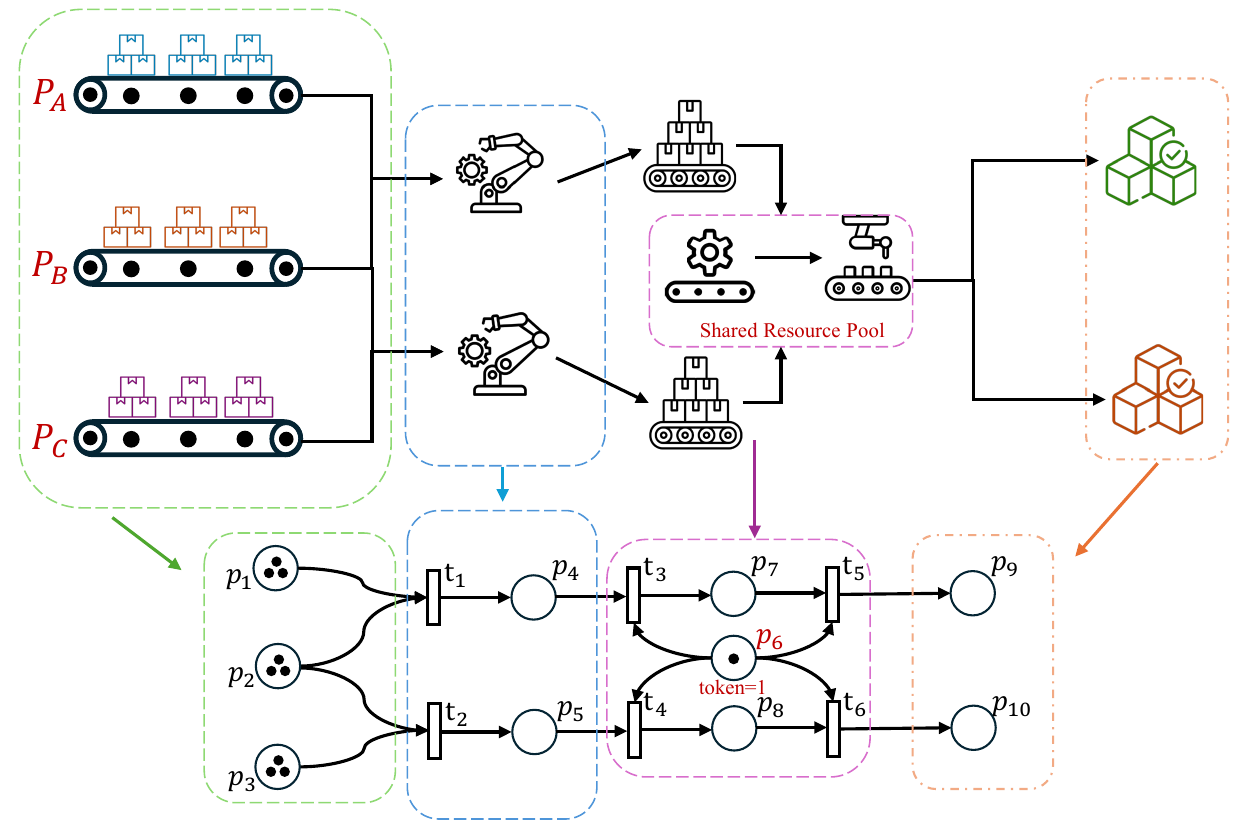}
  \caption{Overview of the industrial case study and the associated Petri net model. The upper panel depicts the production flow with three inbound lines, parallel processing routes, and a capacity-limited shared resource pool. The lower panel presents the corresponding Petri net.}
  \label{fig2}
\end{figure}

The proposed method is evaluated on a real-world inspired multi-product automated manufacturing cell. As illustrated in Fig.~\ref{fig2}, the cell receives workpieces from three heterogeneous inbound lines, denoted as $P_A$, $P_B$, and $P_C$. After entering the cell, jobs proceed through multiple processing operations organized in parallel production routes. The processed items are then merged into a downstream workstation that represents a shared and capacity-limited resource pool. This station cannot simultaneously serve the demands of different product types, which induces contention and requires explicit sequencing decisions. After completing this shared-resource stage, the system outputs two distinct finished product types.

Fig.~\ref{fig2} presents both a process overview and the corresponding Petri net abstraction. The upper part shows the physical material flow of the manufacturing cell, while the lower part presents the Petri net structure used in the experiments. In the Petri net model, places represent buffers, intermediate states, and resource availability, whereas transitions represent processing events. In particular, the shared workstation is modeled by a dedicated resource place with a single token (capacity 1), which enforces mutual exclusion when different job streams request the same bottleneck resource. This case study therefore captures key characteristics of industrial discrete-event systems, including concurrency, synchronization, and resource conflicts, making it suitable for assessing Petri net reachability planning under realistic constraints.

\section{Experiments}
\label{sec.E}

The proposed framework is validated using an industrial case study to demonstrate practical effectiveness in a representative setting. Section~\ref{sec.D} details the industrial scenario and the corresponding Petri net model. Based on this model, Section~\ref{sec.E1} defines six benchmark reachability configurations that share the same net structure but differ in initial and target markings, which enables systematic control of problem difficulty. Section~\ref{sec.E2} evaluates the proposed method on these configurations. Finally, Section~\ref{sec.E3} compares performance across multiple LLMs,which assesses robustness under different model implementations.

\subsection{Experimental Setup}
\label{sec.E1}

All experiments are conducted on the same case study Petri net $\mathcal{N}=(P,T,Pre,Post)$ introduced in Section~\ref{sec.C}. Six reachability configurations are constructed by fixing the net structure fixed while varying the initial and target markings $\big(M_0^{(i)}, M_g^{(i)}\big)$ to induce different levels of resource contention and production requirements. The final marking must match $M_g^{(i)}$ on all places in $P$. Places not explicitly specified in $M_g^{(i)}$ are treated as having zero tokens. Table~\ref{Table1} summarizes the six marking pairs used in the evaluation.

We report four metrics: (1) the overall success rate (\%); (2) the first pass success rate (\%); (3) repair success rate (\%); and (4) average number of retries required until termination.

For each configuration, five independent runs are performed. In each run, the solver executes up to a maximum of 10 attempts. The run terminates immediately once a valid sequence is verified, or stops when the retry budget is exhausted. Unless otherwise specified, all LLM calls use ChatGPT-4o with temperature set to 0.

\begin{table}[b!]
  \centering
  \caption{Six reachability configurations on the case-study Petri net}
  \label{Table1}

  \scriptsize
  \setlength{\tabcolsep}{3pt}
  \renewcommand{\arraystretch}{1.05}

  \begin{tabularx}{\columnwidth}{c >{\raggedright\arraybackslash}X >{\raggedright\arraybackslash}X}
    \toprule
    Case & Initial Marking $M_0^{(i)}$ & Target Marking $M_g^{(i)}$ \\
    \midrule
    case01 & $P_{1}=4,\, P_{2}=8,\, P_{3}=4,\, P_{6}=1$ &
            $P_{6}=1,\, P_{9}=4,\, P_{10}=4$ \\
    case02 & $P_{1}=4,\, P_{2}=8,\, P_{3}=1,\, P_{6}=1$ &
            $P_{6}=1,\, P_{9}=1,\, P_{10}=4$ \\
    case03 & $P_{1}=14,\, P_{2}=24,\, P_{3}=3,\, P_{6}=3$ &
            $P_{5}=1,\, P_{6}=1,\, P_{8}=2,\, P_{9}=7,\, P_{10}=7$ \\
    case04 & $P_{1}=11,\, P_{2}=22,\, P_{3}=8,\, P_{6}=2$ &
            $P_{5}=2,\, P_{6}=1,\, P_{8}=1,\, P_{9}=8,\, P_{10}=8$ \\
    case05 & $P_{1}=35,\, P_{2}=54,\, P_{3}=6,\, P_{6}=6$ &
            $P_{3}=6,\, P_{5}=5,\, P_{6}=1,\, P_{7}=2,\, P_{8}=3,\, P_{9}=14,\, P_{10}=11$ \\
   case06 & $P_{1}=30,\, P_{2}=40,\, P_{3}=5,\, P_{5}=6,\, P_{6}=8,\, P_{7}=2,\, P_{8}=3,\, P_{9}=0,\, P_{10}=0$ &
\parbox[t]{\linewidth}{%
  $P_{1}=5,\, P_{2}=5,\, P_{3}=7,$\\
  $P_{5}=4,\, P_{6}=1,\, P_{7}=5,$\\
  $P_{8}=7,\, P_{9}=10,\, P_{10}=8$
} \\

    \bottomrule
  \end{tabularx}
\end{table}

\subsection{Results on the Six Case Study Instances}
\label{sec.E2}

Table~\ref{Table2} reports the performance of the proposed method on the six marking configurations of the fixed case study Petri net. Across all 30 runs (5 runs per case), the method achieves a $100\%$ overall success rate, demonstrating that the framework can reliably certify valid firing sequences under the exact marking criterion. On the easier configurations (case01--case04), first pass success remains high, and the average number of retries per run is below 0.4, indicating that the LLM frequently outputs a correct sequence. In contrast, the more challenging configurations exhibit substantially lower first-pass success. In particular, case05 has $0\%$ first-pass success but still reaches $100\%$ overall success through repair, requiring on average 2.40 retries per run. This suggests that the effectiveness of reflection and reptompting strategy. Similarly, case06 benefits from the repair loop, reaching $80\%$ repair success with an average of 0.80 retries per run.

Overall, these results highlight the reflection and reprompting mechanism substantially improves robustness on harder cases, converting initial failures into verified solutions and thereby stabilizing performance.

\begin{table}[htbp]
  \centering
  \caption{Performance on six case study instances}
  \label{Table2}

  \scriptsize
  \resizebox{\columnwidth}{!}{%
  \begin{tabular}{ccccc}
    \toprule
    \textit{Case} &
    \makecell[c]{Overall success \\rate(\%)} &
    \makecell[c]{First pass \\success rate(\%)} &
    \makecell[c]{Repair success \\rate(\%)} &
    \makecell[c]{Avg. retries \\per run} \\
    \midrule
    case01 & 100   & 100   & -     & - \\
    case02 & 100   & 80    & 20    & 0.2 \\
    case03 & 100   & 60    & 40    & 0.4 \\
    case04 & 100   & 80    & 20    & 0.2 \\
    case05 & 100   & 0     & 100   & 2.4 \\
    case06 & 100   & 20    & 80    & 0.8 \\
    \addlinespace[1pt]
    Total  & 100   & 56.67 & 43.33 & 0.67 \\
    \bottomrule
  \end{tabular}%
  }
\end{table}%

\subsection{Comparison Across Different LLMs}
\label{sec.E3}

To examine whether the proposed framework is tied to a specific LLM, we evaluate the same closed-loop pipeline using three representative models: GPT-4o, Deepseek-chat, and Gemini-3-Flash. Table~\ref{Table3} summarizes the aggregated results over the six case study instances. The data presented in the table is the final result obtained by each LLM after running five times on six examples.

Overall, the proposed method consistently achieves high feasibility across different LLMs.
Both GPT-4o and Deepseek-chat reach a $100\%$ overall success rate, while Gemini-3-Flash attains $96.67\%$, indicating that the framework remains effective even when the underlying generative model changes. Notably, when initial attempts fail, the reflection and reprompting enables the system to recover and produce verified firing sequences in a large portion of runs. This is reflected in repair success rate. 

While different LLMs exhibit different first pass performance and repair costs, the same verification guided closed loop provides a model-independent mechanism for rejecting invalid steps, diagnosing failures, and iteratively repairing candidate sequences toward the exact target marking.
\begin{table}[h]
  \centering
  \caption{Cross-LLM comparison of the proposed method on six case study instances}
  \label{Table3}

  \scriptsize
  \setlength{\tabcolsep}{1.5pt}
  \renewcommand{\arraystretch}{0.95}

  \begin{tabular*}{\columnwidth}{@{\extracolsep{\fill}} c c c c c @{}}
    \toprule
    \textit{LLMs Model} &
    \makecell[c]{Overall success\\rate (\%)} &
    \makecell[c]{First pass success\\rate (\%)} &
    \makecell[c]{Repair success\\rate (\%)} &
    \makecell[c]{Avg. retries\\per run} \\
    \midrule
    GPT-4o & 100   & 56.67 & 43.33 & 0.67 \\
    \addlinespace[1pt]
    Deepseek-V3 (chat) & 100   & 66.67 & 33.33 & 0.33 \\
    \addlinespace[1pt]
    Gemini-3-Flash  & 96.67 & 70.00  & 26.67 & 1.13 \\
    \bottomrule
  \end{tabular*}%
\end{table}%

\section{Conclusion and Future Work}
\label{sec.F}

This paper studies Petri-net reachability planning and firing sequence under shared resource constraints. We propose an LLM-based solving framework augmented with a two stage reflection and reprompting mechanism to improve correctness and robustness. Unlike one shot sequence generation with an LLM, our framework first prompts the model to produce a candidate firing sequence and then verifies it via step by step replay, checking transition enabledness at each marking and, upon completion, determining whether the resulting marking matches the target specification. When verification fails, the framework drives the LLM to diagnose the failure and formulate a repair strategy, and then regenerates a revised sequence accordingly. This closed-loop process mitigates the inefficiency and uncertainty of naive repeated trials.

On an industrial inspired, multi product automated manufacturing cell case study, we fix the Petri net structure and construct six benchmark configurations with distinct initial and target markings, covering varying degrees of resource contention and different combinations of production requirements. The experimental results indicate that the proposed approach can effectively solve this class of Petri nets, ensuring that the LLM outputs executable and valid firing sequences, thereby yielding feasible schedules for the manufacturing cell. These results further demonstrate the feasibility and practical value of our framework for discrete event manufacturing systems where concurrency, synchronization, and shared resource conflicts coexist.

Regarding computational cost, the total runtime is approximately
the number of LLM calls times the per-call inference latency, since step-by-step
verification is linear in the sequence length and comparatively negligible. The
number of calls is bounded by the retry budget and stays small in our experiments,
so the wall-clock time is governed mainly by the chosen model and its serving
infrastructure rather than by the framework itself.

Future work can proceed along three directions. First, the framework can be extended to larger scale and more complex industrial settings such as multi-objective optimization, dynamic order insertion, and flexible production. Second, the proposed approach can be integrated with complementary methods, including reinforcement learning, machine learning, and heuristic optimization algorithm. Third, the inference cost can be reduced through stronger first-pass prompting, smaller distilled models, or caching.

\bibliographystyle{IEEEtran}
\bibliography{reference}

\end{document}